\documentclass[aps,prl,twocolumn,groupedaddress,showpacs]{revtex4}

\usepackage{graphicx}

\begin{document}


\title{Non-linear Elastic Response in Solid Helium: critical velocity or strain?}


\author{James Day}
\author{Oleksandr Syshchenko}
\author{John Beamish}
\affiliation{Department of Physics, University of Alberta,
Edmonton, Alberta, Canada, T6G 2G7}


\date{\today}

\begin{abstract}
Torsional oscillator experiments show evidence of mass decoupling in
solid $^4$He.  This decoupling is amplitude dependent, suggesting a
critical velocity for supersolidity.  We observe similar behavior
in the elastic shear modulus.  By measuring the shear modulus over
a wide frequency range, we can distinguish between an amplitude
dependence which depends on velocity and one which depends on some
other parameter like displacement.  In contrast to the torsional
oscillator behavior, the modulus depends on the magnitude of stress,
not velocity.   We interpret our results in terms of the motion of
dislocations which are weakly pinned by $^3$He impurities but which
break away when large stresses are applied.
\end{abstract}


\maketitle



Helium is a uniquely quantum solid and recent torsional oscillator (TO)
measurements\cite{Kim04-1941, Kim06-115302, Rittner07-175302, Kondo07-695,
Penzev07-677, Aoki07-015301, Keiderling09-032040, Hunt09-632}
provide evidence for a supersolid phase in hcp $^4$He.  At temperatures below about
200 mK the TO frequency increases, suggesting that some of the $^4$He decouples from
the oscillator.  This evidence of "non-classical rotational inertia" (NCRI) inspired searches
for other unusual thermal or mechanical  behavior in solid helium.  Heat capacity
measurements\cite{Lin07-449, Lin09-125302} do show a small peak near the onset
temperature of decoupling, supporting the idea of a phase transition in solid $^4$He.
Mass flow is an obvious possible signature of supersolidity but experiments in this
temperature range\cite{Day06-105304, Sasaki06-1098, Sasaki07-205302} show no
pressure-induced flow through the solid (although recent measurements\cite{Ray08-235301, Ray09-224302}
at higher temperatures showed intriguing behavior).

We recently made low frequency measurements\cite{Day07-853,
Day09-214524} of the shear modulus of hcp $^4$He and found a large stiffening, with the same temperature dependence
as the frequency changes seen in TO experiments.  This modulus increase also had the
same dependence on measurement amplitude and on $^3$He concentration and, like the
TO decoupling, its onset was accompanied by a dissipation peak.  It is clear that the shear
stiffening and the TO decoupling are closely related.  Subsequent experiments\cite{West09-598}
with $^3$He showed similar elastic stiffening in the hcp phase below 0.4 K, but
not in the bcc phase (the bcc phase exists only over a
temperature range in $^4$He).  TO measurements with hcp $^3$He did not, however, show any
sign of a transition in the temperature range where the stiffening occurred, nor
was a transition seen with bcc $^3$He.  The stiffening appears to depend on crystal
structure (appearing in the hcp but not the bcc phase) while the TO frequency and
dissipation changes occur only for the bose solid, $^4$He.

Although it is clear that solid $^4$He shows unusual behavior below 200 mK, the
interpretation in terms of supersolidity rests almost entirely upon torsional
oscillator experiments.  In addition to frequency changes which imply mass decoupling,
two other features of the TO experiments are invoked as evidence of superflow.
One is the "blocked annulus" experiment\cite{Kim04-1941, Rittner08-155301} in which
NCRI is greatly reduced by inserting a barrier into the flow path, thus indicating
that long-range coherent flow is involved.  The other is the reduction of the
NCRI fraction when the oscillation amplitude exceeds a critical value\cite{Kim04-1941}.
In analogy to superfluidity in liquid helium, this is interpreted
in terms of a critical velocity v$_c$ (of order 10 $\mu$m/s), above which flow becomes
dissipative.  However, torsional oscillators are resonant devices and measurements
made at a single frequency cannot distinguish an amplitude dependence which sets
in at a critical velocity from one which begins at a critical displacement or a
critical acceleration.  A recent experiment\cite{Aoki07-015301} used a compound
torsional oscillator which operated in two modes, allowing measurements to be made
on the same solid $^4$He sample at two different frequencies (496 and  1173 Hz).
The amplitude dependence scaled somewhat better with velocity than with
acceleration or displacement, supporting the superflow interpretation of TO
experiments.  However, recent measurements\cite{Kojima09-private} in which one
mode was driven at large amplitude while monitoring
the low-amplitude response of the other mode gave unexpected results.  The
suppression of NCRI, as seen in the low amplitude mode, appeared to depend on
the acceleration generated by the high amplitude mode, rather than on the velocity.
To settle the question of whether the amplitude dependence seen in torsional
oscillators reflects a critical velocity for superflow, measurements over a
wider frequency range are needed.

We have made direct measurements\cite{Day07-853} of the amplitude dependence of the shear modulus,
$\mu$, of hcp $^4$He in a narrow gap of thickness D.  An AC voltage, V, with
angular frequency $\omega$ = 2$\pi$f, is applied to a shear transducer
(with piezoelectric coefficient d$_{15}$) to generate a displacement $\delta$x=d$_{15}V$
at its surface.  This produces a quasi-static shear strain in the helium, $\epsilon$ = $\delta$x/D.
The resulting stress, $\sigma$, generates a charge, q, and thus a current, I = $\omega$q,
in a second transducer on the opposite side of the gap.  The shear modulus
$\mu$=$\sigma$/$\epsilon$ is then proportional to I/fV.   In contrast to torsional
oscillator measurements, this technique is non-resonant and so we could measure the shear modulus
over a wide frequency range, from a few Hz (a limit set by preamplifier noise)
to a maximum of  2500 Hz (due to interference from the first acoustic resonance in our cell,
around 8 kHz).    The technique is very sensitive, allowing us to make modulus measurements
at strains as low as $\epsilon$ $\approx$ 10$^{-9}$, corresponding to stresses $\sigma$ $\approx$ 0.02 Pa.
The drive voltage can be increased substantially without heating the sample, so measurements
can be made at strains up to  $\epsilon$ $\approx$ 10$^{-5}$, allowing us to study the amplitude
dependence, including hysteretic effects, at low temperatures.

Figure 1 shows the temperature dependence of the normalized shear modulus $\mu$/$\mu$$_o$
for an hcp $^4$He sample at a pressure of 38 bar and a frequency of 2000 Hz.  The crystal
was grown from standard isotopic purity $^4$He (containing about 0.3 ppm $^3$He) using the
blocked capillary method.  A piezoelectric shear stack\cite{piceramic}  was used to generate strains
in a narrow gap (D = 500 $\mu$m) between it and a detecting transducer.  Other
experimental details are the same as in refs. 16 to 18.  The curves in Fig. 1 correspond
to different transducer drive voltages, i.e. to different strains, and were measured during
cooling from a temperature of 0.7 K.  The shear modulus increases at low temperature, with
$\Delta$$\mu$/$\mu$$_o$ $\approx$ 17$\%$ at the lowest amplitude.  Similar stiffening was seen in
all hcp $^4$He crystals\cite{Day07-853, Day09-214524}, although the onset temperature
was lower in crystals of higher isotopic purity and the magnitude of the modulus change
varied by a factor of about 2 from sample to sample (the TO NCRI varies much more, by a
factor of 1000,  although its temperature dependence is always  essentially the
same\cite{West09-598}).  The amplitude dependence of the
shear modulus stiffening is essentially the same as that of the TO NCRI.  At the lowest
drive voltages and temperatures, both are independent of amplitude but they decrease at high amplitudes.
The onset of stiffening shifts to lower temperatures at high amplitudes, as does the onset
of TO decoupling.   However, in the case of elastic measurements, it is more natural to
think of this behavior in terms of a critical stress (proportional to the strain, i.e. to
transducer displacement), rather than a critical velocity.

\begin{figure}
\includegraphics[width=\linewidth]{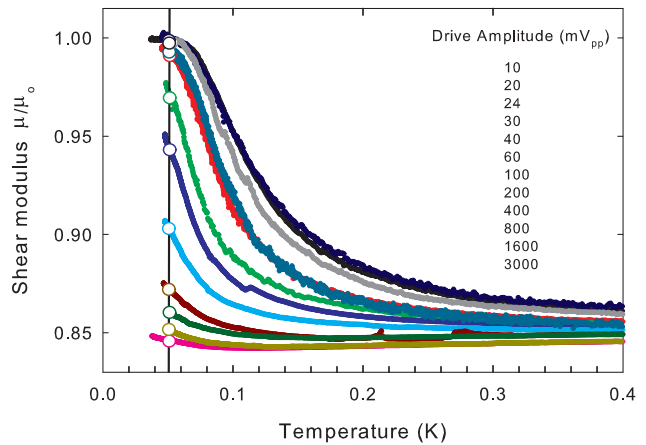}
\caption{Temperature and amplitude dependence of the shear modulus in a solid $^4$He sample
at 38 bar, grown by the blocked capillary method.  The modulus was measured at 2000 Hz
for transducer drive voltages (peak to peak) from 10 mV to 3 V.  Values are normalized
by $\mu$$_o$ the low temperature value at the lowest drive voltage.}
\label{fig:Figure1.EPS}
\end{figure}

The amplitude dependence of the shear modulus is shown in more detail in Fig. 2.  Open
circles show the modulus at 48 mK, taken from the temperature sweeps (i.e. the points marked
by open circles in Fig. 1).  The solid circles are the modulus measured when the drive voltage
was reduced at fixed temperature (48 mK), after cooling from high temperature at the highest
drive amplitude (3 V$_{pp}$).  Figure 2 also shows the corresponding amplitude dependence at
temperatures well above the shear modulus anomaly (open squares are data at 700 mK from
the temperature sweeps of Fig. 1; solid squares are from amplitude sweeps at 800 mK).  The
data taken with the two protocols agree very well.  The critical drive voltage (where the
modulus becomes amplitude dependent) is around 30 mV$_{pp}$ (corresponding to $\sigma$ $\approx$ 4x10$^{-8}$,
$\sigma$ $\approx$ 0.8 Pa) and at the highest drive levels (3 V$_{pp}$ corresponding to $\sigma$ $\approx$ 80 Pa)
the stiffening is almost completely suppressed.  The amplitude dependence of the modulus closely
resembles that of the NCRI in TO experiments, behavior which is attributed to a superflow
critical velocity (typically around 10 $\mu$m/s).  However, even in TO measurements there are
inertial stresses in the helium, produced by its acceleration, which could lead to the observed amplitude
dependence.

\begin{figure}
\includegraphics[width=\linewidth]{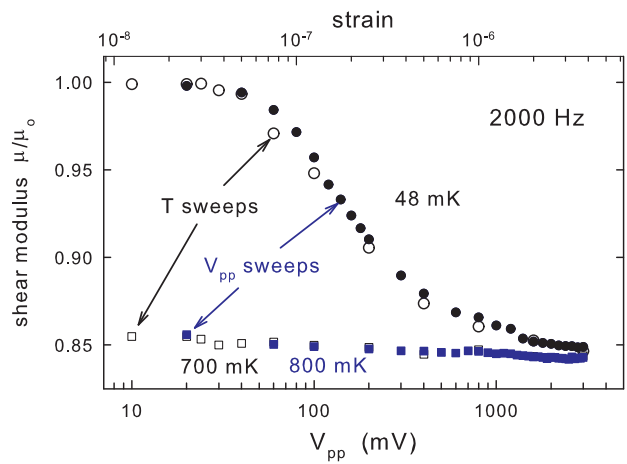}
\caption{Low and high temperature shear modulus for the crystal of Fig. 1,  as a function
of drive voltage (bottom axis) or strain (top axis).  The open symbols are taken from
temperature sweeps at different drive levels (the data of Fig. 1).  The solid symbols are
taken while decreasing the drive voltage at fixed temperature.}
\label{fig:Figure2.EPS}
\end{figure}

A number of experiments\cite{Aoki07-015301, Rittner08-155301, Shimizu09, Clark08-184531}
have shown that the TO amplitude dependence is hysteretic at low
temperatures.   If a sample is cooled at high oscillation amplitude, the apparent NCRI is small.
When the amplitude is reduced at low temperature, the TO frequency (NCRI) rises, becoming
constant below some critical amplitude.  When the drive is then increased at low temperature,
the NCRI does not begin to decrease at the critical amplitude - it remains essentially constant
at substantially larger drives.  This hysteresis between data taken while decreasing and
increasing the drive amplitude disappears at temperatures above about 70 mK.

Figure 3 shows the corresponding hysteresis in the shear modulus.  At 120 mK the maximum stiffening is
about half as large as at 36 mK and already depends on amplitude at the lowest strains shown. The modulus
measured when the amplitude is reduced (open circles) and when it is
subsequently increased (solid circles) agree.  Hysteresis appears when the sample is cooled below
60 mK and is nearly temperature independent below 45 mK.  Figure 3 shows this hysteresis at 36 mK.  The sample was  cooled from high
temperature while driving at high amplitude (3 V$_{pp}$).  The amplitude was
then lowered at 36 mK (open circles) which resulted in a shear modulus increase $\Delta$$\mu$/$\mu$$_o$
of about 15$\%$.  When the amplitude was then raised (solid circles), the modulus remained
constant to much higher amplitude, the same behavior seen for the NCRI in
TO experiments.  At very high amplitudes (above about 1 V$_{pp}$, corresponding to $\epsilon \approx$ 10$^{-6}$,
$\sigma \approx$ 20 Pa) the modulus decreased, nearly closing the hysteresis loop. After each
change we waited 2 minutes for the modulus to stabilize at the new amplitude. The only region where we observed further time
dependence was while increasing the amplitude at drive levels above 1 V$_{pp}$.  The modulus decrease was
sharper when we waited longer at each point.  In acoustic resonance measurements\cite{Day09-214524}, we found that even larger stresses
($\sigma \approx$700 Pa) produced irreversible changes which only disappeared after annealing above 0.5 K.

\begin{figure}
\includegraphics[width=\linewidth]{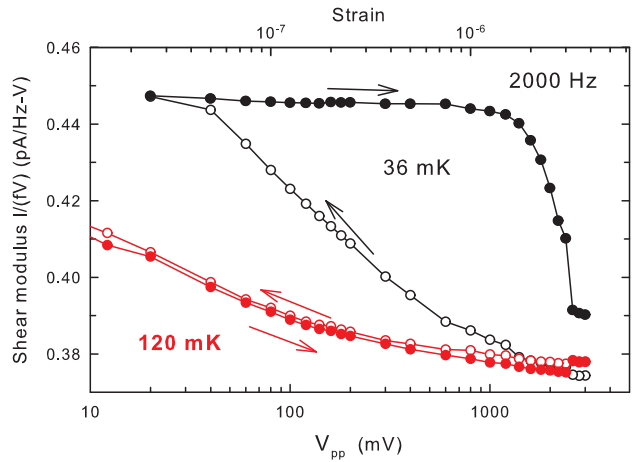}
\caption{Hysteresis between the shear modulus measured while decreasing and while increasing
the drive amplitude at low temperature (36 mK). At 120 mK the amplitude dependence is reversible.}
\label{fig:Figure3.EPS}
\end{figure}

The only obvious mechanism which can produce shear modulus changes as large as those shown in
Figs. 1 to 3 involves the motion of dislocations\cite{Nowick72}.  At low temperatures,
dislocations are pinned by $^3$He impurities and the intrinsic modulus is measured\cite{Day09-214524}.
As the temperature is raised, $^3$He impurities thermally unbind from the dislocations,
allowing them to move and reducing the modulus.  At high amplitudes, elastic stresses can
also tear the dislocations away from the impurities. The hysteresis seen in Fig. 3 can be understood if, when a crystal
is cooled at high strain amplitudes, the rapid motion of dislocations prevents $^3$He atoms
from attaching to them.  When the drive is reduced at low temperatures, impurities can
bind, thus pinning the dislocations and increasing the modulus. Once the $^3$He impurities
bind, the pinning length of dislocations is smaller and larger stresses are required to
unpin them so the modulus retains its intrinsic value to much higher amplitudes.  The critical amplitude for this
"stress-induced breakaway" can be estimated\cite{Iwasa80-1722, Paalanen81-664}
if the dislocation length and impurity binding energy are known.  In single crystals of helium, a typical dislocation
network pinning length\cite{Iwasa79-1119} is (L $\sim$ 5 $\mu$m), and dislocations would
break away from a $^3$He impurity at strain $\epsilon \approx$3x10$^{-7}$. Our crystals are expected to have
higher dislocation densities and smaller network lengths, so breakaway would occur at higher strains
as the amplitude is increased.  Measurements with different $^3$He concentrations would be useful to
confirm that the amplitude dependence is due to this mechanism.

This interpretation of the shear modulus behavior involves elastic stress (which is proportional
to strain) rather than velocity, and so is at odds with the interpretation of the TO amplitude
dependence in terms of a superfluid-like critical velocity.  Since we can make modulus measurements
over a wide frequency range, we can unambiguously distinguish between an amplitude dependence
which scales with stress (or strain) and one which depends on velocity.  Figure 4 shows the modulus
at 18 mK, measured at three different frequencies (2000, 200 and 20 Hz) as the drive amplitude
was reduced from its maximum value.   In Fig. 4a the modulus is plotted vs. shear strain $\epsilon$
(calibrated using the low temperature piezoelectric coefficient of the shear stack, d$_{15}$=1.25 nm/V)
and in Fig. 4b the same data is plotted versus the corresponding velocities (v=$\omega$$\epsilon$D).
The amplitude dependence scales much better with strain than with velocity (and the scaling
with acceleration is even less satisfactory).  The critical strain appears be slightly larger
at lower frequency.

\begin{figure}
\includegraphics[width=\linewidth]{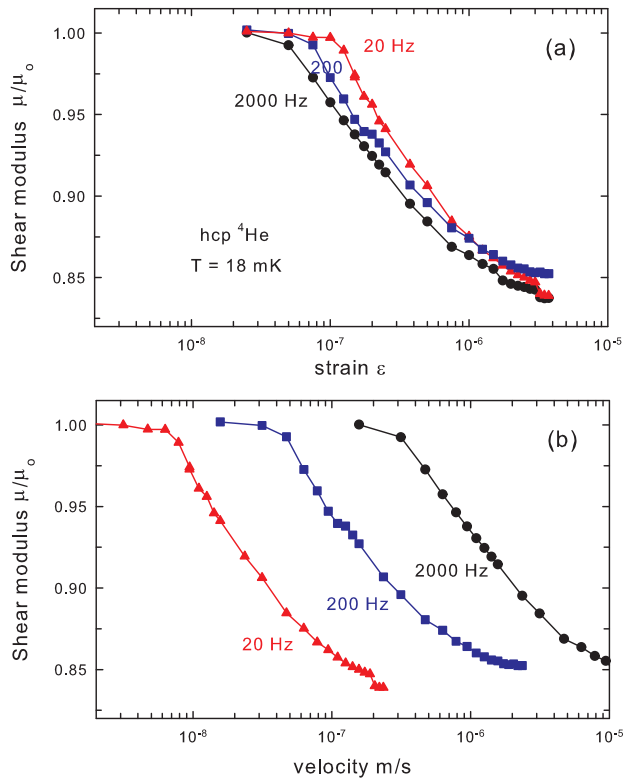}
\caption{Scaling of the shear modulus' amplitude dependence with (a) strain and (b) velocity
for three different frequencies: 20 Hz (triangles), 200 Hz (squares) and 2000 Hz (circles).}
\label{fig:Figure4.EPS}
\end{figure}

It is clear that the amplitude dependence of the shear modulus is most closely associated with
stress (or strain) amplitude, rather than with a superfluid-like critical velocity.  The
many similarities to the TO behavior (e.g. the dependence on temperature, $^3$He concentration
and frequency, the amplitude dependence and its hysteresis) suggest that the apparent velocity
dependence of the NCRI may have a similar origin, e.g. in inertial stresses which exceed the
critical value for the shear modulus.   However, estimates of the inertial stress corresponding
to TO critical velocities give values significantly lower than the critical stress for the shear modulus.
For an annular TO geometry, the maximum inertial stress can be estimated as
$\sigma$=$\rho$t$\omega$v/2, where $\rho$ is the helium density, t is the width of the annular
channel, $\omega$ is the angular frequency of the TO and v is the oscillation velocity.   For
the oscillator of ref. 1, we estimate $\sigma \approx$  0.002 to 0.015 Pa at the apparent critical
velocities of 5 to 38 $\mu$m/s.  Measurements in an open cylindrical TO\cite{Clark08-184531} show amplitude
dependence at velocities corresponding to inertial stresses below 0.01 Pa, also much smaller
than the stresses at which we observe amplitude dependence in the shear modulus.  There is also
other evidence that the TO frequency changes and dissipation are not just mechanical consequences
of the modulus changes.  First, the apparent NCRI is too large to be explained simply by
mechanical stiffening of the torsional oscillator\cite{West09-598, Clark08-184531}.  Secondly,
comparable modulus changes in hcp $^3$He are not reflected in the corresponding TO
behavior\cite{West09-598}.   The existence of a critical velocity for superflow in solid helium
can only be definitively shown if TO measurements can be made over a wide range of
frequency and/or TO geometries.

This work was supported by the Natural Sciences and Engineering
Research Council of Canada.

\bibliography{nonlinear}

\begin{thebibliography}{27}
\expandafter\ifx\csname natexlab\endcsname\relax\def\natexlab#1{#1}\fi
\expandafter\ifx\csname bibnamefont\endcsname\relax
  \def\bibnamefont#1{#1}\fi
\expandafter\ifx\csname bibfnamefont\endcsname\relax
  \def\bibfnamefont#1{#1}\fi
\expandafter\ifx\csname citenamefont\endcsname\relax
  \def\citenamefont#1{#1}\fi
\expandafter\ifx\csname url\endcsname\relax
  \def\url#1{\texttt{#1}}\fi
\expandafter\ifx\csname urlprefix\endcsname\relax\def\urlprefix{URL }\fi
\providecommand{\bibinfo}[2]{#2}
\providecommand{\eprint}[2][]{\url{#2}}

\bibitem[{\citenamefont{Kim and Chan}(2004)}]{Kim04-1941}
\bibinfo{author}{\bibfnamefont{E.}~\bibnamefont{Kim}} \bibnamefont{and}
  \bibinfo{author}{\bibfnamefont{M.~H.~W.} \bibnamefont{Chan}},
  \bibinfo{journal}{Science} \textbf{\bibinfo{volume}{305}},
  \bibinfo{pages}{1941} (\bibinfo{year}{2004}).

\bibitem[{\citenamefont{Kim and Chan}(2006)}]{Kim06-115302}
\bibinfo{author}{\bibfnamefont{E.}~\bibnamefont{Kim}} \bibnamefont{and}
  \bibinfo{author}{\bibfnamefont{M.~H.~W.} \bibnamefont{Chan}},
  \bibinfo{journal}{Phys. Rev. Lett.} \textbf{\bibinfo{volume}{97}},
  \bibinfo{pages}{115302} (\bibinfo{year}{2006}).

\bibitem[{\citenamefont{Rittner and Reppy}(2007)}]{Rittner07-175302}
\bibinfo{author}{\bibfnamefont{A.~S.~C.} \bibnamefont{Rittner}}
  \bibnamefont{and} \bibinfo{author}{\bibfnamefont{J.~D.} \bibnamefont{Reppy}},
  \bibinfo{journal}{Phys. Rev. Lett.} \textbf{\bibinfo{volume}{98}},
  \bibinfo{pages}{175302} (\bibinfo{year}{2007}).

\bibitem[{\citenamefont{Kondo et~al.}(2007)\citenamefont{Kondo, Takada,
  Shibayama, and Shirahama}}]{Kondo07-695}
\bibinfo{author}{\bibfnamefont{M.}~\bibnamefont{Kondo}},
  \bibinfo{author}{\bibfnamefont{S.}~\bibnamefont{Takada}},
  \bibinfo{author}{\bibfnamefont{Y.}~\bibnamefont{Shibayama}},
  \bibnamefont{and}
  \bibinfo{author}{\bibfnamefont{K.}~\bibnamefont{Shirahama}},
  \bibinfo{journal}{J. Low Temp. Phys.} \textbf{\bibinfo{volume}{148}},
  \bibinfo{pages}{695} (\bibinfo{year}{2007}).

\bibitem[{\citenamefont{Penzev et~al.}(2007)\citenamefont{Penzev, Yasuta, and
  Kubota}}]{Penzev07-677}
\bibinfo{author}{\bibfnamefont{A.}~\bibnamefont{Penzev}},
  \bibinfo{author}{\bibfnamefont{Y.}~\bibnamefont{Yasuta}}, \bibnamefont{and}
  \bibinfo{author}{\bibfnamefont{M.}~\bibnamefont{Kubota}},
  \bibinfo{journal}{J. Low Temp. Phys.} \textbf{\bibinfo{volume}{148}},
  \bibinfo{pages}{677} (\bibinfo{year}{2007}).

\bibitem[{\citenamefont{Aoki et~al.}(2007)\citenamefont{Aoki, Graves, and
  Kojima}}]{Aoki07-015301}
\bibinfo{author}{\bibfnamefont{Y.}~\bibnamefont{Aoki}},
  \bibinfo{author}{\bibfnamefont{J.}~\bibnamefont{Graves}}, \bibnamefont{and}
  \bibinfo{author}{\bibfnamefont{H.}~\bibnamefont{Kojima}},
  \bibinfo{journal}{Phys. Rev. Lett.} \textbf{\bibinfo{volume}{99}},
  \bibinfo{pages}{015301} (\bibinfo{year}{2007}).

\bibitem[{\citenamefont{Keiderling et~al.}(2009)\citenamefont{Keiderling, Aoki,
  and Kojima}}]{Keiderling09-032040}
\bibinfo{author}{\bibfnamefont{M.}~\bibnamefont{Keiderling}},
  \bibinfo{author}{\bibfnamefont{Y.}~\bibnamefont{Aoki}}, \bibnamefont{and}
  \bibinfo{author}{\bibfnamefont{H.}~\bibnamefont{Kojima}},
  \bibinfo{journal}{J. Phys. Conf. Ser.} \textbf{\bibinfo{volume}{150}},
  \bibinfo{pages}{032040} (\bibinfo{year}{2009}).

\bibitem[{\citenamefont{Hunt et~al.}(2009)\citenamefont{Hunt, Pratt, Gadagkar,
  Yamashita, Balatsky, and Davis}}]{Hunt09-632}
\bibinfo{author}{\bibfnamefont{B.}~\bibnamefont{Hunt}},
  \bibinfo{author}{\bibfnamefont{E.}~\bibnamefont{Pratt}},
  \bibinfo{author}{\bibfnamefont{V.}~\bibnamefont{Gadagkar}},
  \bibinfo{author}{\bibfnamefont{M.}~\bibnamefont{Yamashita}},
  \bibinfo{author}{\bibfnamefont{A.~V.} \bibnamefont{Balatsky}},
  \bibnamefont{and} \bibinfo{author}{\bibfnamefont{J.~C.} \bibnamefont{Davis}},
  \bibinfo{journal}{Science} \textbf{\bibinfo{volume}{324}},
  \bibinfo{pages}{632} (\bibinfo{year}{2009}).

\bibitem[{\citenamefont{Lin et~al.}(2007)\citenamefont{Lin, Clark, and
  Chan}}]{Lin07-449}
\bibinfo{author}{\bibfnamefont{X.}~\bibnamefont{Lin}},
  \bibinfo{author}{\bibfnamefont{A.}~\bibnamefont{Clark}}, \bibnamefont{and}
  \bibinfo{author}{\bibfnamefont{M.~H.~W.} \bibnamefont{Chan}},
  \bibinfo{journal}{Nature} \textbf{\bibinfo{volume}{449}},
  \bibinfo{pages}{1025} (\bibinfo{year}{2007}).

\bibitem[{\citenamefont{Lin et~al.}(2009)\citenamefont{Lin, Clark, Cheng, and
  Chan}}]{Lin09-125302}
\bibinfo{author}{\bibfnamefont{X.}~\bibnamefont{Lin}},
  \bibinfo{author}{\bibfnamefont{A.~C.} \bibnamefont{Clark}},
  \bibinfo{author}{\bibfnamefont{Z.~G.} \bibnamefont{Cheng}}, \bibnamefont{and}
  \bibinfo{author}{\bibfnamefont{M.~H.~W.} \bibnamefont{Chan}},
  \bibinfo{journal}{Phys. Rev. Lett.} \textbf{\bibinfo{volume}{102}},
  \bibinfo{pages}{125302} (\bibinfo{year}{2009}).

\bibitem[{\citenamefont{Day and Beamish}(2006)}]{Day06-105304}
\bibinfo{author}{\bibfnamefont{J.}~\bibnamefont{Day}} \bibnamefont{and}
  \bibinfo{author}{\bibfnamefont{J.}~\bibnamefont{Beamish}},
  \bibinfo{journal}{Phys. Rev. Lett.} \textbf{\bibinfo{volume}{96}},
  \bibinfo{pages}{105304} (\bibinfo{year}{2006}).

\bibitem[{\citenamefont{Sasaki et~al.}(2006)\citenamefont{Sasaki, Ishiguru,
  Caupin, Maris, and Balibar}}]{Sasaki06-1098}
\bibinfo{author}{\bibfnamefont{S.}~\bibnamefont{Sasaki}},
  \bibinfo{author}{\bibfnamefont{R.}~\bibnamefont{Ishiguru}},
  \bibinfo{author}{\bibfnamefont{F.}~\bibnamefont{Caupin}},
  \bibinfo{author}{\bibfnamefont{H.~J.} \bibnamefont{Maris}}, \bibnamefont{and}
  \bibinfo{author}{\bibfnamefont{S.}~\bibnamefont{Balibar}},
  \bibinfo{journal}{Science} \textbf{\bibinfo{volume}{313}},
  \bibinfo{pages}{1098} (\bibinfo{year}{2006}).

\bibitem[{\citenamefont{Sasaki et~al.}(2007)\citenamefont{Sasaki, Caupin, and
  Balibar}}]{Sasaki07-205302}
\bibinfo{author}{\bibfnamefont{S.}~\bibnamefont{Sasaki}},
  \bibinfo{author}{\bibfnamefont{F.}~\bibnamefont{Caupin}}, \bibnamefont{and}
  \bibinfo{author}{\bibfnamefont{S.}~\bibnamefont{Balibar}},
  \bibinfo{journal}{Phys. Rev. Lett.} \textbf{\bibinfo{volume}{99}},
  \bibinfo{pages}{205302} (\bibinfo{year}{2007}).

\bibitem[{\citenamefont{Ray and Hallock}(2008)}]{Ray08-235301}
\bibinfo{author}{\bibfnamefont{M.~W.} \bibnamefont{Ray}} \bibnamefont{and}
  \bibinfo{author}{\bibfnamefont{R.~W.} \bibnamefont{Hallock}},
  \bibinfo{journal}{Phys. Rev. Lett.} \textbf{\bibinfo{volume}{100}},
  \bibinfo{pages}{235301} (\bibinfo{year}{2008}).

\bibitem[{\citenamefont{Ray and Hallock}(2009)}]{Ray09-224302}
\bibinfo{author}{\bibfnamefont{M.~W.} \bibnamefont{Ray}} \bibnamefont{and}
  \bibinfo{author}{\bibfnamefont{R.~W.} \bibnamefont{Hallock}},
  \bibinfo{journal}{Phys. Rev. B} \textbf{\bibinfo{volume}{79}},
  \bibinfo{pages}{224302} (\bibinfo{year}{2009}).

\bibitem[{\citenamefont{Day and Beamish}(2007)}]{Day07-853}
\bibinfo{author}{\bibfnamefont{J.}~\bibnamefont{Day}} \bibnamefont{and}
  \bibinfo{author}{\bibfnamefont{J.}~\bibnamefont{Beamish}},
  \bibinfo{journal}{Nature} \textbf{\bibinfo{volume}{450}},
  \bibinfo{pages}{853} (\bibinfo{year}{2007}).

\bibitem[{\citenamefont{Day et~al.}(2009)\citenamefont{Day, Syshchenko, and
  Beamish}}]{Day09-214524}
\bibinfo{author}{\bibfnamefont{J.}~\bibnamefont{Day}},
  \bibinfo{author}{\bibfnamefont{O.}~\bibnamefont{Syshchenko}},
  \bibnamefont{and} \bibinfo{author}{\bibfnamefont{J.}~\bibnamefont{Beamish}},
  \bibinfo{journal}{Phys. Rev. B} \textbf{\bibinfo{volume}{79}},
  \bibinfo{pages}{214524} (\bibinfo{year}{2009}).

\bibitem[{\citenamefont{West et~al.}(2009)\citenamefont{West, Syshchenko,
  Beamish, and Chan}}]{West09-598}
\bibinfo{author}{\bibfnamefont{J.~T.} \bibnamefont{West}},
  \bibinfo{author}{\bibfnamefont{O.}~\bibnamefont{Syshchenko}},
  \bibinfo{author}{\bibfnamefont{J.}~\bibnamefont{Beamish}}, \bibnamefont{and}
  \bibinfo{author}{\bibfnamefont{M.~H.~W.} \bibnamefont{Chan}},
  \bibinfo{journal}{Nature Physics} \textbf{\bibinfo{volume}{5}},
  \bibinfo{pages}{598} (\bibinfo{year}{2009}).

\bibitem[{\citenamefont{Rittner and Reppy}(2008)}]{Rittner08-155301}
\bibinfo{author}{\bibfnamefont{A.~S.~C.} \bibnamefont{Rittner}}
  \bibnamefont{and} \bibinfo{author}{\bibfnamefont{J.~D.} \bibnamefont{Reppy}},
  \bibinfo{journal}{Phys. Rev. Lett.} \textbf{\bibinfo{volume}{101}},
  \bibinfo{pages}{155301} (\bibinfo{year}{2008}).

\bibitem[{Koj()}]{Kojima09-private}
\bibinfo{note}{Kojima, H. (private communication).}

\bibitem[{pic()}]{piceramic}
\bibinfo{note}{Http://www.piceramic.com/ Model 141-03}.

\bibitem[{\citenamefont{Shimizu et~al.}(2009)\citenamefont{Shimizu, Yasuta, and
  Kubota}}]{Shimizu09}
\bibinfo{author}{\bibfnamefont{N.}~\bibnamefont{Shimizu}},
  \bibinfo{author}{\bibfnamefont{Y.}~\bibnamefont{Yasuta}}, \bibnamefont{and}
  \bibinfo{author}{\bibfnamefont{M.}~\bibnamefont{Kubota}},
  \emph{\bibinfo{title}{{\it arXiv:cond-mat/09031326}}} (\bibinfo{year}{2009}).

\bibitem[{\citenamefont{Clark and Chan}(2008)}]{Clark08-184531}
\bibinfo{author}{\bibfnamefont{A.~C.} \bibnamefont{Clark}} \bibnamefont{and}
  \bibinfo{author}{\bibfnamefont{M.~H.~W.} \bibnamefont{Chan}},
  \bibinfo{journal}{Phys. Rev. B} \textbf{\bibinfo{volume}{77}},
  \bibinfo{pages}{184513} (\bibinfo{year}{2008}).

\bibitem[{\citenamefont{Nowick and Berry}(1972)}]{Nowick72}
\bibinfo{author}{\bibfnamefont{A.~S.} \bibnamefont{Nowick}} \bibnamefont{and}
  \bibinfo{author}{\bibfnamefont{B.~S.} \bibnamefont{Berry}},
  \emph{\bibinfo{title}{Anelastic Relaxation in Crystalline Solids}}
  (\bibinfo{publisher}{Academic Press, New York}, \bibinfo{year}{1972}).

\bibitem[{\citenamefont{Iwasa and Suzuki}(1980)}]{Iwasa80-1722}
\bibinfo{author}{\bibfnamefont{I.}~\bibnamefont{Iwasa}} \bibnamefont{and}
  \bibinfo{author}{\bibfnamefont{H.}~\bibnamefont{Suzuki}},
  \bibinfo{journal}{J. Phys. Soc. Japan} \textbf{\bibinfo{volume}{49}},
  \bibinfo{pages}{1722} (\bibinfo{year}{1980}).

\bibitem[{\citenamefont{Paalanen et~al.}(1981)\citenamefont{Paalanen, Bishop,
  and Dail}}]{Paalanen81-664}
\bibinfo{author}{\bibfnamefont{M.~A.} \bibnamefont{Paalanen}},
  \bibinfo{author}{\bibfnamefont{D.~J.} \bibnamefont{Bishop}},
  \bibnamefont{and} \bibinfo{author}{\bibfnamefont{H.~W.} \bibnamefont{Dail}},
  \bibinfo{journal}{Phys. Rev. Lett.} \textbf{\bibinfo{volume}{46}},
  \bibinfo{pages}{664} (\bibinfo{year}{1981}).

\bibitem[{\citenamefont{Iwasa and Suzuki}(1979)}]{Iwasa79-1119}
\bibinfo{author}{\bibfnamefont{I.}~\bibnamefont{Iwasa}} \bibnamefont{and}
  \bibinfo{author}{\bibfnamefont{H.}~\bibnamefont{Suzuki}},
  \bibinfo{journal}{J. Phys. Soc. Japan} \textbf{\bibinfo{volume}{46}},
  \bibinfo{pages}{1119} (\bibinfo{year}{1979}).

\end{thebibliography}

\end{document}